\documentclass[twocolumn,aps,prb,superscriptaddress,a4paper,showpacs,showkeys]{revtex4-1}

\usepackage{graphicx}
\usepackage{dcolumn}
\usepackage{bm}
\usepackage{type1cm}
\usepackage[normalem]{ulem}

\usepackage{amssymb,amsmath}

\usepackage[dvips]{color}

\newcommand{\mm}[1]{\mbox{$#1$}}

\newcommand{\istd}{\mathrm{i}}
\newcommand{\estd}{\mathrm{e}}

\newcommand{\mbf}[1]{\bm{#1}}

\newcommand{\rsms}{real-space multiple-scattering}

\newcommand{\crcu}{\mbox{(creat)$_{2}$CuCl$_{4}$}}
\newcommand{\crCu}{\crcu}

\newcommand{\ea}{{\it et al.}}

\begin{document}

\title{Including atomic vibrations in XANES calculations:
  polarization-dependent damping of the fine structure at the Cu $K$
  edge of \crcu}



\author{O. \surname{\v{S}ipr}}
\email{sipr@fzu.cz}
\homepage{http://www.fzu.cz/~sipr} 
\affiliation{Institute of Physics of the ASCR v.~v.~i.,
  Cukrovarnick\'{a}~10, CZ-162~53~Prague, Czech Republic }

\author{J. \surname{Vack\'{a}\v{r}}} 
\affiliation{Institute of Physics of the ASCR v.~v.~i.,
  Cukrovarnick\'{a}~10, CZ-162~53~Prague, Czech Republic }

\author{A. \surname{Kuzmin}} 
\affiliation{Institute of Solid State Physics, University of
  Latvia,Kengaraga street 8, LV-1063 Riga, Latvia}

\date{\today}

\begin{abstract}
Atomic vibrations are usually not taken into account when analyzing
x-ray absorption near edge structure (XANES) spectra.  One of the
reasons is that including the vibrations in a formally exact way is
quite complicated while the effect of vibrations is supposed to be
small in the XANES region.  By analyzing polarized Cu $K$ edge x-ray
absorption spectra of creatinium tetrachlorocuprate [\crCu], we
demonstrate that a technically simple method, consisting in
calculating the XANES via the same formula as for static systems but
with a modified free-electron propagator which accounts for
fluctuations of interatomic distances, may substantially help in
understanding XANES of some layered systems.  In particular we show
that the difference in the damping of the x-ray absorption fine
structure oscillations for different polarisations of the incoming
x-rays cannot be reproduced by calculations which rely on a static
lattice but it can be described if atomic vibrations are accounted for
in such a way that individual creatinium and CuCl$_{4}$ molecular
blocks are treated as semi-rigid entities while the mutual positions
of these blocks are subject to large mean relative displacements.
\end{abstract}

\pacs{78.70.Dm}

\keywords{XANES, vibrations, multiple-scattering formalism}

\maketitle


\section{Introduction}

\label{sec-intro}

The importance of atomic vibrations for a proper description of x-ray
absorption spectra (XAS) in
the extended x-ray absorption fine structure (EXAFS) region has been
recognised since the early years of this technique.\cite{BP+76,Greegor1979} The
vibrations have to be accounted for to obtain a proper damping of
x-ray absorption fine structure (XAFS)
oscillations at high energies.  Including the vibrations in XAS
calculations is in general a difficult task because in principle one
has to evaluate the spectrum for all possible geometric configurations
(distributions of instantaneous positions of atoms) and to average the
signals with proper weights.  Doing this explicitly might be
computationally very demanding so usually some further simplifications
are invoked, such as assumption of a Gaussian or nearly-Gaussian
distribution regarding the atomic positions and of validity of
single-scattering and plane-wave approximations regarding the effect
of vibrations on the photoelectron state.\cite{BP+76,LCE+81}
Sometimes, an explicit averaging of the XAFS signal has been employed
as well --- both when treating a thermal
disorder\cite{BNF+89,BCW+07,Kuzmin2009,Anspoks2012} 
and when treating a static disorder.\cite{BNF+89,SDR04,KDF06,Anspoks2012}

So far the attention has been focused mostly either on the pre-edge
region or on the EXAFS region.  The interest in the pre-edge region is
driven mainly by the effort to explain the emergence of some pre-peaks
as a consequence of a vibrational local symmetry breaking which allows
some otherwise forbidden dipole
transitions.\cite{VKN+98,DBK+10,MCB+12} The interest in the EXAFS
region stems from the fact that accounting for vibrations is
obligatory for an accurate fitting of EXAFS signal to extract reliable
values of coordination numbers, bond lengths and mean-square relative
displacements (MSRD's).\cite{Greegor1979,Anspoks2012} The x-ray
near-edge structure (XANES) region covering photoelectron energies up
to about 50~eV and the intermediate energy region (photoelectron
energies 50--100~eV) have not attracted much attention in this
respect.  One of the reasons is that one expects the effect of
vibrations to be small at low $k$-values close to the absorption edge,
as it is reflected by the Debye-Waller factor $\exp(-2\sigma^2k^2)$
which occurs in the conventional EXAFS formula.\cite{LCE+81} Besides,
the use of the Fourier filtering procedure affects the quality of the
EXAFS signal at low $k$-values, so that this part is usually excluded
from the structural analysis.  

However, there might be effects in the XANES or intermediate energy
regions for which the vibrations could be important, and it would be
useful to check it.  Analysis of specific examples would certainly be
quite instructive in this respect.  An appealing option is to focus on
polarised spectra of layered systems, because (i) polarised XANES
contains significantly more information than orientationally-averaged
spectra, meaning also that polarised XANES presents a much more
stringent test to the theory\cite{BKD01,SS01,CRG+12} and (ii) for
layered systems is it more likely that the vibrations will damp
different scattering signals in a different way, giving thus rise to
features in the spectra which might not be possible to explain unless
these vibrations are taken into account.

Including Gaussian or harmonic vibrations into EXAFS
calculations is a relatively straightforward procedure, as long as we
deal with energy region where single-scattering and plane-wave
approximation can be used.  Here, the main challenge is to find the
vibrational modes of the system and to use them for obtaining the
appropriate parameters {\em ab
  initio}.\cite{DB+98,PR+99,VRR+07,DMB+09}

Including vibrations into XANES calculations, where
multiple-scattering effects have to be considered, is more difficult
because, among others, in this case one should deal with displacements
of more than just two atoms at a time.  One possible direction is to
employ Taylor expansion around equilibrium positions to obtain usable
formula for the statistical average of the signals for different
geometric configurations.  This approach was pursued by Benfatto
\ea\cite{BNF+89} who started with analytic formula describing signals
of order $n$ and applied the expansion regarding both the
free-electron propagator and the scattering phase shifts.  A similar
route was taken by Fujikawa \ea\cite{FRW+99} who also applied Taylor
expansion but this time regarding only the propagator.

Another possible approach is to rely on scattering path expansion to
employ Debye-Waller damping factors specific for each scattering
path.\cite{PR+99} Similarly to the scheme of Benfatto
\ea,\cite{BNF+89} this scheme is more suitable to the EXAFS than to
the XANES because it can be used only if the scattering path expansion
converges.\cite{RA00}

Recently, the problem of calculating electronic structure for a system
of vibrating atoms was formulated in the framework of alloy theory, by
identifying each instantaneous atomic position with a new atomic type
and treating the problem within the coherent potential approximation
(CPA).\cite{MKW+13} Due to its general nature, this scheme could be
applied to XAS.  The CPA framework is quite powerful as it implicitly
includes the effect of vibrations both on the propagator and on the
phase shifts. On the other hand, because all atomic displacements are
treated independently, this scheme cannot describe a situation where
some atoms vibrate only a little relatively to each other but at the
same time vibrate quite a lot relatively to other atoms, i.e., when
both strongly correlated and uncorrelated atomic vibrations are
present at the same time.

All these procedures have various degree of accuracy and
sophistication and proved to be useful in various
circumstances.\cite{LP+96,HKF+99,HFN+03,EMK+11,MKW+13} At the same
time, their implementation is not simple.  It is noteworthy in this
respect that even though methodological works undoubtedly present a
significant progress,\cite{BNF+89,DB+98,FRW+99,PR+99,VRR+07,DMB+09}
practical applications of these approaches to XANES calculations
remain to be quite rare.  A broader use of these techniques could be
helped by demonstrating that useful results can be obtained even by
simple implementations of the procedures mentioned above.  If these
techniques are to be used in the XANES region, one has to opt for a
formulation which treats the multiple scattering exactly.  On top of
that, it would be convenient if correlations between movements of
various atoms was taken into account to some extent.  In this work we
focus on testing a simplified version of the method of Fujikawa
\ea,\cite{FRW+99} which consists in calculating XANES by using
essentially the same formula as for static systems, just with a
modified free-electron propagator to account for pair-wise
fluctuations of interatomic distances.  Effectively this means that
the multiple scattering is treated exactly while the effect of
vibrations is treated within the plane wave approximation.  Respective
formulas were already incorporated earlier, e.g., in the {\sc feff9}
code.\cite{RKV+10,feff-code} However, to the best of our knowledge
they have not really been used for solving practical problems of XANES
analysis so far.

We suggest that by means of the procedure outlined in
Sec.~\ref{sec-vibr} one can describe experimentally observable trends
that could not be described within the static lattice model.  We
demonstrate this on the case of Cu $K$ edge spectra of \crcu.  Copper
$K$ edge XAS of Cu-containing complexes was subject to intensive
research in the past.  There has been a still unsettled debate about
the ``local and many-body'' or ``delocalised and one-electron''
interpretation of x-ray near-edge structure (XANES) at the Cu $K$ edge
in systems containing CuCl$_{n}$ or CuO$_{n}$
blocks.\cite{HSH+82,KYA+84,SPB+85,YKK+86,KKT+89,GEG+90,BLC+91,BKD01,SS01,Kos+02,CMM+06,CRG+12}
An important place in these discussion was given to the angular or
polarisation dependence of the spectra close to the edge. However, not
much attention has been paid to the polarisation dependence of XAFS
further above the edge, for photoelectron energies 30--100~eV.  A
striking feature in this energy region is a big difference in how the
XAFS oscillations are damped with the increasing energy for different
polarisations of the incoming x-rays.  We demonstrate in this work
that these differences cannot be explained by calculations which rely
on a static lattice.  By including the vibrations in XANES
calculations we show that the observed anisotropy of XAFS damping can
be reproduced if the mean square relative displacements are chosen so
that individual molecular blocks in the \crCu\ crystal are treated as
semi-rigid entities while the mutual positions of these blocks are
subject to large disorder.  Application of the procedure thus offers a
view on the nature of vibrations in the material in question.  It is
especially suited for analysing XAFS of layered systems.


\section{Methods}


\subsection{Treatment of vibrations}

\label{sec-vibr}

We start by discussing an approach to account for atomic vibrations
within the real space multiple-scattering formalism.  Let us recall
that for a static system, the XANES transition rate can be evaluated
as\cite{Vve92,NBD03,RA00}
\begin{equation}
\mu(\hbar\omega) \, = \,
-\frac{ 2\pi^{2} m^{2}}{\hbar^{5} k^{2}} \,  \Im
 \sum_{L L'} M^{\ast}_{L}
\, \tau^{00}_{L L'} \, M_{L'}
\;\; ,
\label{eqbas}
\end{equation}
where $k=\sqrt{2m(\hbar\omega-E_{0})/\hbar^{2}}$ is the photoelectron
wave vector, $M_{L}$ is the atomic-like transition matrix element and
$\tau^{00}_{L L'}$ is the scattering-path operator comprising all the
scattering events,
\begin{equation}
\tau^{00}_{L L'} \: = \:
t^{0}_{L} \delta_{L L'} \, + \, \sum_{p} \sum_{L''}
t^{0}_{L} \, G^{0p}_{L L''} \, \tau^{p0}_{L'' L'}
\;\;.
\label{eqtau}
\end{equation}
The definitions of the single-site scattering matrix $t^{p}_{L}$ and
of the free-electron propagator $G^{pq}_{L L''}$ can be found in the
above mentioned reviews.\cite{Vve92,NBD03,RA00} The angular-momentum
subscript $L$ is an abbreviation for the pair ($\ell$,$m$) and the sum
$\sum_{p}$ runs over all atoms in the cluster, with $0$ denoting the
photoabsorbing site.

The method we use for accounting for the vibrations is essentially the
same as what was introduced by Fujikawa \ea\cite{FRW+99} as
``renormalisation of the scattering series by a plane wave thermal
factor''.  As mentioned in Sec.~\ref{sec-intro}, the formula was
included also in the {\sc feff9} code\cite{RKV+10,feff-code} but no
discussion of it was given.  The procedure consists in calculating the
XANES using same formula as for static systems but replacing the
free-electron propagator $G^{pq}_{L L''}$ in Eq.~(\ref{eqtau})
according to
\begin{equation}
G^{pq}_{L L''} \rightarrow
G^{pq}_{L L''} \, \estd^{-k^{2} \sigma_{pq}^{2}}
\;\;,
\label{eqsubs}
\end{equation}
where $\sigma_{pq}^{2}$ is the MSRD characterising the fluctuations of
the distance between the atoms $p$ and $q$.

Fujikawa \ea\cite{FRW+99} derived their formula by expanding the
free-electron propagator --- Eq.~(\ref{eqsubs}) above is equivalent to
Eq.~(3.18) in Ref.~\onlinecite{FRW+99}.  (Note, however, that these
authors used a more accurate and at the same time more complicated
expression in their own application of this
technique.)\cite{HKF+99,HFN+03} It is worthy to note that the {\em
  ansatz} Eq.~(\ref{eqsubs}) can be obtained also by another way, by
means of exploiting the analogy with EXAFS formula.  To show this, we
start with the expression for EXAFS oscillations for static
systems:\cite{LCE+81}
\begin{align}
\chi \, &\equiv \,
\frac{\mu-\mu_{0}}{\mu_{0}}
   \nonumber \\
 &= \,
\sum_{p}
\frac{ 3 (\mbf{\hat{\varepsilon}} \cdot \mbf{\hat{R}}_{p})^{2} }
{kR^{2}_{p}} \,
\Im \left[ f_{p}(k) \,
\estd^{2 \istd k R_{p} + 2 \istd \delta^{0}_{\ell=1}} \right]
\;\;.
\label{eqexafs}
\end{align}
Here $\mbf{\varepsilon}$ is the polarisation vector, $R_{p}$ is the
distance between the photoabsorbing atom and the atom $p$, $f_{p}(k)$
is the backward scattering amplitude, and $\delta^{0}_{\ell=1}$ is the
scattering phase shift of the central atom (${\ell=1}$ in case of the $K$ edge).
For a vibrating system one can extend Eq.~(\ref{eqexafs}) to the
well-known formula\cite{BP+76,LCE+81}
\begin{equation}
\chi \, \, = \,
\sum_{p}
\frac{ 3 (\mbf{\hat{\varepsilon}} \cdot \mbf{\hat{R}}_{p})^{2} }
{kR^{2}_{p}} \,
\Im \left[ f_{p}(k) \,
\estd^{2 \istd k R^{0}_{p} + 2 \istd \delta^{0}_{\ell=1}} \,
\estd^{-2  k^{2} \sigma_{p}^{2}}
\right]
\label{eqstd}
\;\;.
\end{equation}
In deriving Eq.~(\ref{eqstd}) from Eq.~(\ref{eqexafs}), one has to
neglect the influence of vibrations on the electronic structure
represented by $f_{p}(k)$ and $\delta^{0}_{\ell=1}$ and assume that
the vibrations are harmonic.
There is, however, yet another way to arrive at Eq.~(\ref{eqstd}).
Namely, one can start with the basic expression (\ref{eqbas}), perform
the substitution (\ref{eqsubs}) and only afterwards apply all the
approximations which transform Eq.~(\ref{eqbas}) into
Eq.~(\ref{eqexafs}), namely, restrict the multiple scattering just to
single scattering and apply the plane wave approximation to the free
electron propagator.  

Having this in mind, we can see that if one calculates XANES spectrum
along the same procedures as in the static case with only performing
the substitution (\ref{eqsubs}), one gets XANES spectrum for a
vibrating system, with the scattering events treated exactly and with
the effects of vibrations treated approximatively.  In particular, (i)
the effect of vibrations is restricted to changes in the geometry,
i.e., the scattering phase shifts are the same as for a static system,
(ii) the effect of vibrations is restricted to varying the bond
lengths for each pair independently, neglecting thus, among others,
bond angles variations and (iii) the effect of vibrations is treated
within the plane wave approximation (while the electron scattering
itself and hence the bulk of the spectral shape are calculated
exactly, i.e., with curved waves).  It should be mentioned that
correlations between movements of different atomic pairs can be
included only partially, by a suitable choice of the MSRD for each
pair, as it is done in Sec.~\ref{sec-results} below.

We will show in the following that despite all the approximations the
{\em ansatz} Eq.~(\ref{eqsubs}) can describe the effect of vibrations
so that features seen in the experiment can be interpreted as arising
from vibrations.

\subsection{Technical details of XANES calculations}

\begin{figure}
\includegraphics[viewport=0 0 558 408,height=55mm]{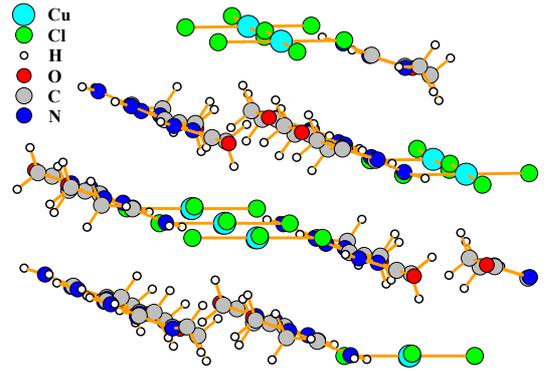}%
\caption{Structural diagram of \crcu.  The $z$ axis is
  perpendicular to the CuCl$_{4}$ blocks.}
\label{fig-geom}
\end{figure}

We apply the method outlined above to the Cu $K$ edge XANES of \crCu.
The crystal structure of \crCu\ is monoclinic (crystal group 14).
Lattice vectors lengths we use are $a$=8.106~\AA, $b$=7.835~\AA,
$c$=13.685~\AA, angles between the lattice vectors are
$\alpha$=90$^{\circ}$, $\beta$=114$^{\circ}$, $\gamma$=90$^{\circ}$.
A structural diagram is shown in Fig.~\ref{fig-geom}.  Most of the
calculations presented here were performed within the real space
multiple-scattering formalism,\cite{NBD03} using the {\sc rsms} code.
\cite{rsms,VSP86} We used a non-self-consistent muffin-tin potential
constructed according to the Mattheiss prescription (superposition of
potentials and charge densities of isolated atoms).  To alleviate the
deficiencies of the muffin-tin approximation, some calculations were
done also while inserting additional ``empty spheres'' into the
interstitial region.  The positions and radii of these empty spheres
were obtained via the {\sc xband} code,\cite{xband} relying on
procedures commonly used in the linear muffin-tin orbital (LMTO)
method calculations.\cite{AAT+98} The core hole left by the excited
photoelectron was treated within the final state approximation
(``relaxed and screened model'').\cite{SS01,Vve92,NBD03} The exchange
and correlation effects were accounted for via an energy-independent
$X\alpha$\ potential with the Kohn-Sham value of $\alpha$=0.67.  We do
not employ energy-dependent exchange-correlation potential because
there is no universal recipe which one to choose for a particular
case.\cite{CQ+95,Ank99,HC+07} Our focus will be on the anisotropy of
XAFS damping and selecting a different exchange-correlation potential
would just shift the positions of all the peaks no matter to which
polarization they belong, hence this is not really important for our
study.

The XANES spectra presented in this work were obtained for clusters of
radii of 7.8~\AA\ (185 atoms including H atoms, 103 atoms disregarding
H atoms). Fairly well size-converged spectra were obtained already for
clusters of radii of 4.5~\AA\ (33 atoms including H atoms, 19 atoms
disregarding H atoms). Constructing the muffin-tin potential for
hydrogen atoms has been problematic sometimes.\cite{WTA+00} However,
in our case this is not an issue because we checked that the spectra
calculated with H atoms included and with H atoms ignored are
practically identical.

Raw spectra were broadened by an energy-dependent Lorentzian to
simulate the combined effect of the experimental resolution, of the
decay of the core hole and of the decay of the excited photoelectron.
The constant part of the broadening was set to 1.50~eV, the
energy-dependent part of the broadening was set to \mm{0.08E}, where
$E$ is the photoelectron energy.  Such a choice leads to reasonable
spectra.  The agreement between theory and experiment could further be
improved by optimizing the broadening function.\cite{MJW82,BDN+03}
While such a procedure is useful when structural optimization is
sought, in our case it would have no influence on the conclusions
because we focus on the {\em differences} in XAFS damping for
different polarizations while the Lorentzian broadening is isotropic.


\subsection{Plane-waves calculations}

\label{sec-pseudo}

Band-structure calculations relying on plane waves basis set and
pseudopotentials were performed using the {\sc abinit}
code.\cite{GAA+09} First, self-consistent calculation was done using
16 $\mbf{k}$-points in the irreducible Brillouin zone.  Afterwards,
angular-momentum-projected density of states (DOS) within a sphere
around the Cu atom with radius of 1.24~\AA\ was calculated using 38
$\mbf{k}$-points.  The cut-off energy for the plane waves was set to
40~H.  All these parameters were checked for convergence.  The core
hole was ignored.  The results presented here were obtained using a
Hartwigsen-Goedeker-Hutter type pseudopotentials available at the {\sc
  abinit} web-site.\cite{pseudogoed} We verified that very similar
results are obtained also for pseudopotentials of the
Troullier-Martins type.\cite{pseudofhi,pseudoTM}

Because of the use of pseudopotentials, we did not have access to the
core wave functions and, consequently, we were not able to calculate
XANES spectra directly.  However, if the core hole is neglected and
dipole selection rule is invoked, polarised XANES spectra can be taken
as proportional to the appropriate angular-momentum component of the
projected DOS multiplied by a smooth function which corresponds to the
square of the atomic transition matrix element $M_{L}$.  The
proportionality between XAS and DOS is, strictly speaking, true only
under specific symmetry requirements which guarantee that the
scattering path operator \mm{\tau^{00}_{LL'}} entering
Eq.~(\ref{eqbas}) is diagonal.  However, in most cases this is
fulfilled with sufficient accuracy.  We checked explicitly using the
{\sc rsms} code that XAS and DOS are indeed proportional in our case.
Having all this in mind, we obtained XANES spectra via {\sc abinit}
simply by taking the projected DOS and multiplying it by a smooth
sinusoidal-like function which increases monotonously from 1.0 at the
absorption edge to 1.6 at the end of the energy region we cover.  We
chose this particular form of the DOS-to-XAS proportionality function
just for convenience, our conclusions are not dependent on it.
Polarisation-sensitivity of the spectra was achieved by taking the
($\ell$=1,$m$=0) DOS component to get the XANES with the
\mm{\varepsilon \| z}\ polarisation and the ($\ell$=1,$m$=$\pm$1) DOS
component to get the XANES with the \mm{\varepsilon \perp
  z}\ polarisation.  The same convolution of the raw results by an
energy-dependent Lorentzian curve was applied as in case of
\rsms\ calculations.


\section{Application to the Cu $K$ edge XAS of \crcu}

\label{sec-results}

\begin{figure}
\includegraphics[viewport=0.9cm 0.5cm 9.0cm 16.0cm,width=85mm]{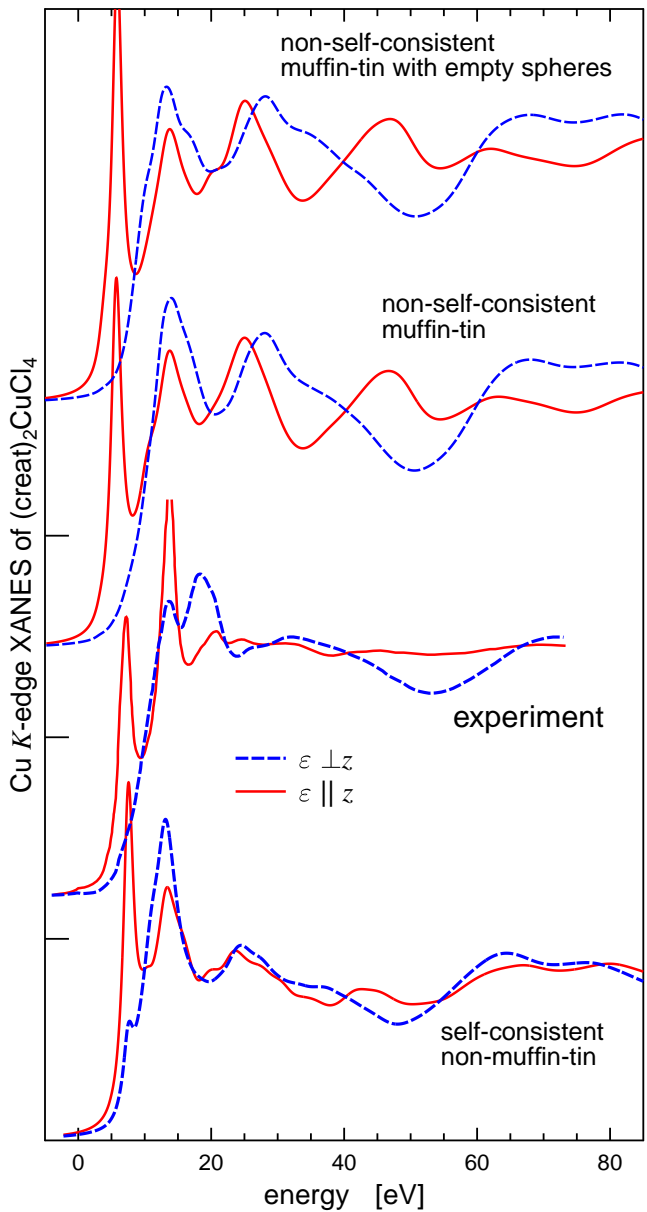}%
\caption{Experimental polarised Cu $K$ edge XANES of
  \crcu\ (Ref.~\onlinecite{KYA+84}) compared to spectra calculated for
  a static crystal.  The two uppermost panels display spectra obtained
  via the {\sc rsms} code, the lowermost panel displays spectra
  obtained via the {\sc abinit} code. The origin of the energy scale
  is arbitrary.}
\label{fig-static}
\end{figure}

In the following we will demonstrate the potential of the procedure
outlined in Sec.~\ref{sec-vibr} by analysing the Cu $K$ edge XAS of
\crcu.  The experimental spectrum taken from the work of Kosugi
\ea\cite{KYA+84} is shown in the second from the bottom panel of
Fig.~\ref{fig-static}.  The polarisation vector of the incoming x-rays
either lies in the plane defined by the photoabsorbing Cu atom and its
four nearest Cl neighbours ($\varepsilon \perp z$) or is perpendicular
to this plane ($\varepsilon \| z$, cf.\ Fig.~\protect\ref{fig-geom}).
Unlike most of previous works on the Cu $K$ edge XAS of \crcu, our
focus is not on the first 10~eV above the edge but on the higher
energy region, in particular on the strong suppression of the fine
structure for the $\varepsilon \| z$ spectrum for energies 20~eV and
more above the edge.


\subsection{Calculations relying on a static lattice}

\label{sec-static}

First we find out whether the anisotropic damping of XAFS oscillations
can be reproduced by calculations done for a static crystal.
Theoretical spectra obtained using the {\sc rsms} code are shown above
the experimental spectrum in Fig.~\ref{fig-static}, spectrum obtained
using the {\sc abinit} code is shown below the experimental spectrum.
Spectra labelled $\varepsilon \| z$ were calculated for linearly
polarised x-rays, spectra labelled $\varepsilon \perp z$ were
calculated for circularly polarised x-rays.  The horizontal alignment
of the spectra was done by hand so that the position of the
peak around 12~eV for $\varepsilon \| z$ matches the experiment.

It is evident that none of the calculations describes the suppression
of the fine structure for $\varepsilon \| z$.  In particular, this is
true for calculations which use a non-self-consistent potential (the
upper two graphs in Fig.~\ref{fig-static}, {\sc rsms} code) as well as
for calculations which use a self-consistent potential (the lowermost
graph in Fig.~\ref{fig-static}, {\sc abinit} code).  Likewise, similar
results are obtained when relying on the muffin-tin approximation (the
second graph from the top in Fig.~\ref{fig-static}), when the
muffin-tin approximation is improved via use of additional empty
spheres (the uppermost graph) and when the muffin-tin approximation is
not used altogether (the lowermost graph).  The static effect of the
1$s$ core hole as described within the final state approximation is
practically negligible --- we found that the spectra obtained via the
{\sc rsms} code for a potential with the core hole can be hardly
distinguished from spectra obtained for a ground state potential (not
shown here).  It is thus evident that the anisotropic damping of XAFS
oscillations for energies 20~eV or more above the edge cannot be
reproduced by calculations done for a static \crcu\ crystal.

As concerns individual spectra, our calculations are unable to
reproduce correctly the peak intensities within the first 10~eV above
the absorption edge.  We conjecture that this is connected with
many-body effects beyond the local density approximation (LDA).  Let
us recall in this regard the discussion about whether the Cu $K$ edge
XANES of complexes containing linear or square-planar CuCl$_{n}$ units
is dominated by many-body (mostly shake-down) features or whether it
is dominated by hybridisation between states of neighbouring CuCl$_{n}$
units.\cite{KYA+84,SPB+85,YKK+86,Kos+02} Similar arguments are also
put forward in connection with XANES of copper
oxides.\cite{KKT+89,GEG+90,BLC+91,BKD01,SS01,CMM+06,CRG+12} To
contribute to this debate is beyond our scope but we would like to
point out that approaches which handle local many-body effects
accurately via quantum chemistry configuration interaction
calculations are usually unable to consider the interaction between
neighbouring CuCl$_{n}$ units while calculations which include this
long-range interaction treat the many-body effects only in a
mean-field way (via the LDA).  Assessing the relative importance of
both contributions is thus difficult.  The failure of our calculations
to reproduce the experiment accurately down to the edge can be seen as
an evidence that many-body effects are important in this region but we
cannot conclude whether they are dominant or not.

Interestingly, the shape of the broad $\varepsilon \| z$ peak around
20--30~eV cannot be described accurately unless a self-consistent
non-muffin-tin potential is used: the three sub-peaks this feature
consists of are well reproduced by {\sc abinit} calculations but not
by {\sc rsms} calculations.  It appears thus that disregarding the
topic of the anisotropic XAFS damping, the states lying predominantly
in the $xy$ crystallographic plane which contains a lot of atoms are
well-described by the non-self-consistent muffin-tin potential, while
the states in the more-or-less empty interstitial region between the
CuCl$_{4}$-planes can be described properly only if the approximations
laid on the potential are relaxed.  This is plausible --- one expects
that the muffin-tin approximation will be more crude in empty regions
that in regions packed with atoms.

Finally, we should mention that our theoretical spectra do not exhibit
the weak pre-edge structure which appears in the experiment around
zero eV in the scale of Fig.~\ref{fig-static} (x-ray energies about
8978~eV). Our calculations are based on the dipole selection rule,
reflecting thus only states having the $p$ symmetry with respect to
the Cu atoms.  The pre-edge structure, on the other hand, is to a
large extent formed by quadrupole transitions, reflecting thus the $d$
states.  Another contribution to the pre-edge structure may come from
dipole transitions allowed by vibrational symmetry breaking. The role
of quadrupole transitions for the Cu $K$ pre-edge structure was
discussed widely in the past\cite{HSH+82,BKD01,WXH+04} and the same is
true for the effect of vibrations on pre-edge
intensities.\cite{VKN+98,MCB+12} We want to focus rather on the
higher-energy part of the spectrum.


\subsection{Influence of vibrations}

\label{sec-res-vibr}

\begin{figure}
\includegraphics[viewport=0.5cm 0.0cm 17.0cm 18.5cm,scale=0.2]{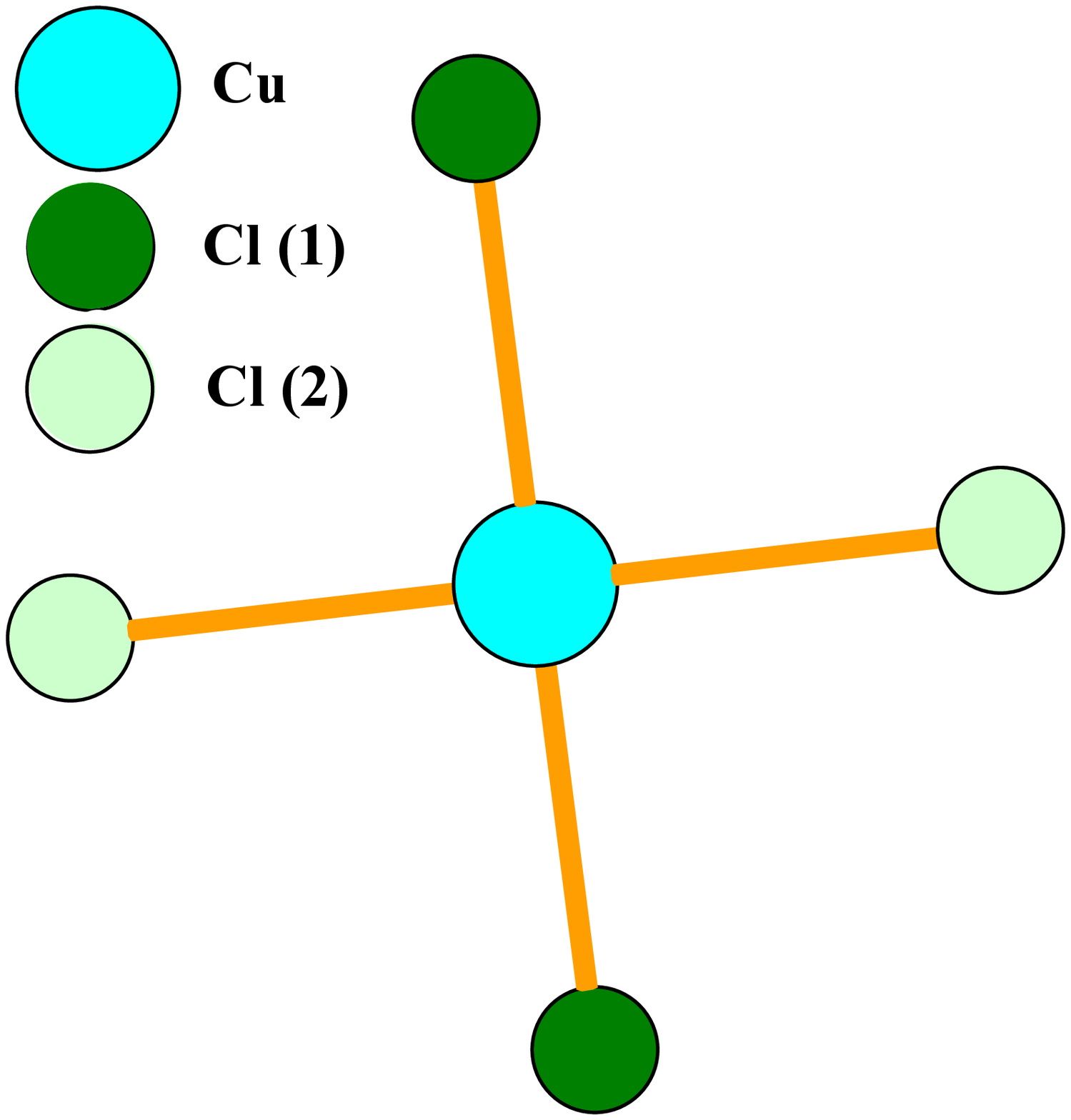}%
\includegraphics[viewport=-7.0cm 0.0cm 17.0cm 18.5cm,scale=0.2]{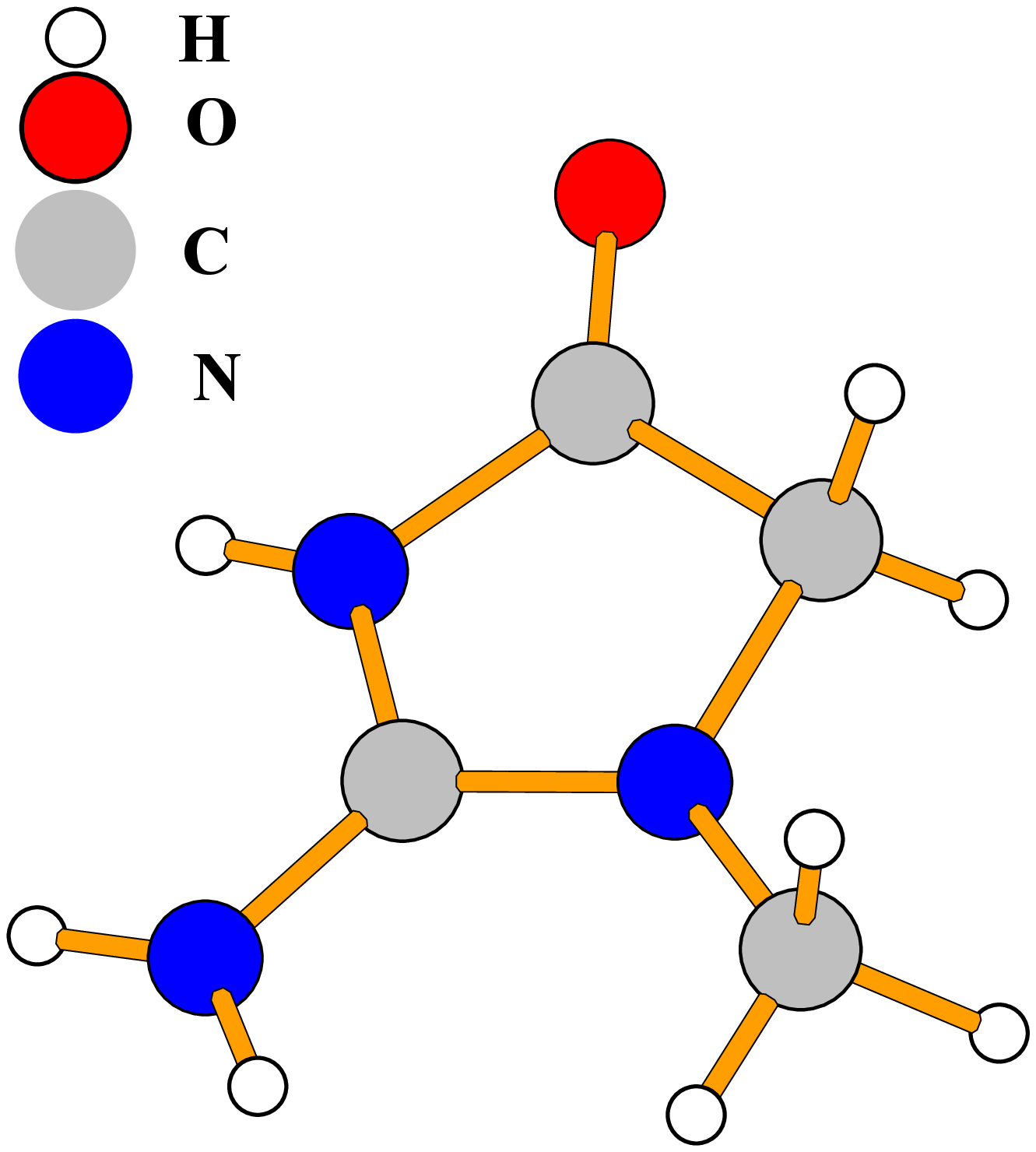}%
\caption{Molecular blocks forming the \crCu\ crystal.}
\label{fig-units}
\end{figure}

Because static lattice calculations were not able to reproduce the
large difference between the damping of XAFS oscillations for the
\mm{\varepsilon \| z}\ and for \mm{\varepsilon \perp
  z}\ polarisations, we performed another set of calculations,
accounting this time for vibrations by the method described in
Sec.~\ref{sec-vibr}.  The occurrence of the anisotropic damping of XAFS in the
experiment suggests that one should consider different MSRD's for
different bonds.  This means that we have to identify which
interatomic distances will be regarded as stiff and which will be
regarded as soft.  By inspecting the \crCu\ structure one can realize
that the \crCu\ crystal is actually formed by creatinium and
CuCl$_{4}$ molecular units which are relatively independent (they are
linked through \mbox{N-H...Cl} hydrogen bonding).\cite{UK+79} These
units are displayed in Fig.~\ref{fig-units}.  We assume that if two
atoms belong to the same molecular unit, their interatomic distance is
characterised by a small MSRD which we denote
$\sigma_{\text{idd}}^{2}$.  If two atoms belong to different molecular
units, their interatomic distance is characterised by a large MSRD,
denoted as $\sigma_{\text{diff}}^{2}$.  At the moment, we have no
clear indication how to chose the values of $\sigma_{\text{idd}}^{2}$
and $\sigma_{\text{diff}}^{2}$, therefore we treat them as model
parameters and set their values in accordance with what was reported
for other systems in the literature.  Our goal is to check whether a
model based on the substitution of Eq.~(\ref{eqsubs}) with two
distinct $\sigma^{2}$ parameters can simulate the anisotropic XAFS
damping which we observe for \crCu.

\begin{figure}
\includegraphics[viewport=0.9cm 1.0cm 9.0cm 15.0cm,width=85mm]{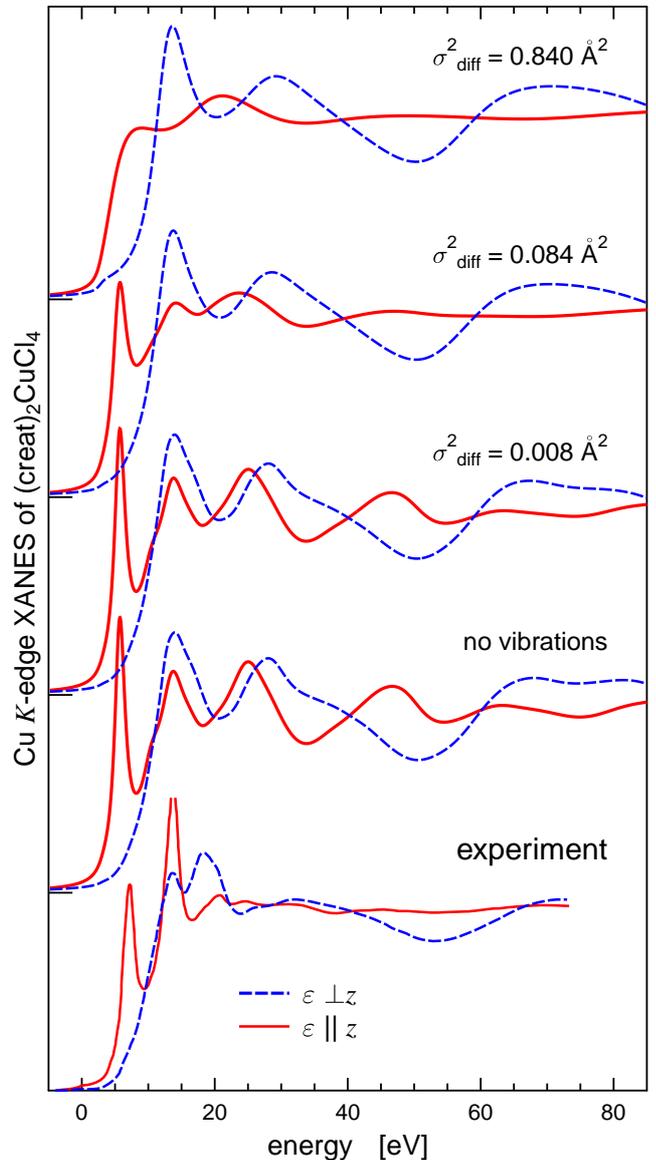}%
\caption{Theoretical Cu $K$ edge XANES of
  \crcu\ calculated while including the vibrational disorder, compared
  to the static lattice calculation and to the experiment.}
\label{fig-vibr}
\end{figure}

We performed a set of calculations with a fixed
$\sigma_{\text{idd}}^{2}$=0.0008~\AA$^{2}$ while varying
$\sigma_{\text{diff}}^{2}$ as 0.008, 0.084, and 0.840~\AA$^{2}$.
These results are compared with results for a static lattice and with
the experimental data (Fig.~\ref{fig-vibr}).  One can see that if we
take $\sigma_{\text{diff}}^{2}$=0.084~\AA$^{2}$, the XAFS oscillations
for the \mm{\varepsilon \| z}\ component are strongly damped for x-ray
energies above 20--30~eV, as required.  A substantially lower
$\sigma_{\text{diff}}^{2}$ does not lead to the desired damping while
a substantially larger $\sigma_{\text{diff}}^{2}$ results in
overdamping of the oscillations at low energies.  We can thus conclude
that our way of including the vibrations in the XANES calculations,
together with the scheme for assigning $\sigma_{\text{idd}}^{2}$ and
$\sigma_{\text{diff}}^{2}$ to atomic pairs depending on whether they
belong to the same molecular unit or not, is a good approximation to
describe the mechanism of the anisotropic damping of XAFS for \crCu.


\section{Discussion}

Our goal was to examine whether a technically simple method to account
for the effect of vibrations in XANES calculation by modifying the
free-electron propagator can be useful in analysing XAFS in certain
systems. By analysing theoretical Cu $K$ edge spectra of
\crCu\ obtained for static and vibrating systems we were able to link
the differences in the damping of experimental XAFS oscillations for
the in-plane and out-of-plane polarisations to the differences in
MSRD's according to whether the respective atoms belong to the same
molecular unit or not.

The MSRD values $\sigma_{\text{idd}}^{2}$=0.0008~\AA$^{2}$ and
$\sigma_{\text{diff}}^{2}$=0.084~\AA$^{2}$ are realistic considering
analogous data obtained for similar systems.  For example, the MSRD
values $\sigma^2$(Cu--Cl)=0.0013-0.0041~\AA$^2$\ in
Ref.~\onlinecite{Tanimizu2007} and 0.008~\AA$^2$\ in
Ref.~\onlinecite{Ivashkevich2008} were obtained for the four-fold Cl
coordination of copper in chloro-complexes.  Note that in these two
cases copper atoms are additionally bonded to two oxygen or nitrogen
atoms, located at shorter distances than four chlorine atoms. As a
result, the Cu--Cl bonding in chloro-complexes is weaker than the
Cu--Cl bonding in \crcu, and, consequently, the corresponding MSRD's
are expected to be larger in chloro-complexes than in \crcu.

We can also make a comparison with crystallographic data.\cite{UK+79}
However, one has to bear in mind that in crystallography one deals
with disorder in the distribution of atomic displacements measured
from a lattice point while in XAS one deals with disorder in the
distribution of interatomic distances.  Correlations between movements
of atoms are responsible for the differences between crystallographic
$\sigma^{2}_{\text{cryst}}$ and XAS-related $\sigma^{2}_{\text{XAS}}$.
Crystallographic data suggest that the vibrations of all atoms in
\crCu\ are strongly anisotropic: $\sigma^{2}_{\text{cryst}}$ along the
$z$ axis is up to four times larger than $\sigma^{2}_{\text{cryst}}$
within to the $xy$ plane.\cite{UK+79} This is consistent with our
picture, because small ``intra-block'' vibrations associated with
$\sigma_{\text{idd}}^{2}$ will be mostly in the $xy$ plane while large
``inter-block'' vibrations associated with $\sigma_{\text{diff}}^{2}$
will be both in the $xy$ plane and along the $z$ axis.  If the
correlations are neglected, an estimate for pair-wise
$\sigma^{2}_{\text{XAS}}$ can be obtained by summing the
crystallographic $\sigma^{2}_{\text{cryst}}$ values for the respective
atom pair.\cite{UK+79} If we do this for the Cu--X pairs along the $z$
axis, where only weak correlations can be expected, we get
$\sigma^{2}_{\text{XAS}}$=0.10--0.17~\AA$^{2}$.  This is comparable
with our $\sigma_{\text{diff}}^{2}$=0.084~\AA$^{2}$.  For vibrations
in the $xy$ plane, $\sigma^{2}_{\text{cryst}}$ is about
0.05--0.09~\AA$^{2}$ if no correlations are assumed.  This is
significantly larger than $\sigma_{\text{idd}}^{2}$ we use.  However,
our assumption of the stiffness of individual molecular blocks means
that there should be large correlations between movements of
respective atoms, so crystallographic $\sigma^{2}_{\text{cryst}}$
cannot be directly compared to our $\sigma_{\text{idd}}^{2}$ in this
respect.

The scheme we employed treats the multiple-scattering in an exact way
but the vibrations in an approximative way.  However, we do not expect this to
have any significant influence on our conclusions.  As concerns the
plane wave approximation laid on the vibrations, it is true that
Fujikawa \ea\cite{FRW+99} found a difference in XANES calculated when
the vibrations were treated within the spherical wave approximation
[Eq.~(3.16) in Ref.~\onlinecite{FRW+99}] and when the vibrations were
treated within the plane wave approximation [Eq.~(3.18) {\em ibid.}]
but this was only for short bond lengths and very large
$\sigma^{2}$=0.4~\AA.  For longer distances, this effect was already
very small.  In our situation, short bond lengths (between atoms of
the same molecular unit) are characterised by small $\sigma^{2}$,
while large $\sigma^{2}$ is associated only with large bond lengths.
Hence, the plane wave approximation should be sufficient for the
vibrations in this case.  As concerns the effect of the vibrations on
the scattering phase shifts, the situation is more complicated ---
earlier studies found some effects in this respect, especially as
concerns the higher-order scattering signals.\cite{BNF+89,YO+96} On
the other hand, we do not expect this effect to dominate in our
situation, because we found that several different potentials lead to
similar results in the energy region where the effect of vibrations
are visible (see Fig.~\ref{fig-static} and the discussion about the
pseudopotentials in Sec.~\ref{sec-pseudo} and about the negligibility of
the core hole effect in Sec.~\ref{sec-static}).

It should be noted that our focus is on the very fact that vibrations
give rise to significant observable effects in the
polarisation-dependence of \crCu\ XANES spectra and not on the
particular values of $\sigma_{\text{idd}}^{2}$ and
$\sigma_{\text{diff}}^{2}$.  The approximations involved in the model
do not allow for taking our $\sigma_{\text{idd}}^{2}$ and
$\sigma_{\text{diff}}^{2}$ values as true best fits.  It should be
also noted that our splitting of the atomic pairs into the
$\sigma_{\text{idd}}^{2}$ and $\sigma_{\text{diff}}^{2}$ classes is
not unique.  E.g., the $\sigma_{\text{idd}}^{2}$ and
$\sigma_{\text{diff}}^{2}$ MSRD's could be ascribed to the pairs
depending on whether the respective atoms belong to the same layer or
not (the layers are depicted figure~\ref{fig-geom}).  We did the
calculations also for this model and found that the results are almost
the same as for the molecular-units-based model.  This is not so
surprising because atoms belonging to the same unit belong also to the
same layer.  We also verified that the particular value of
$\sigma_{\text{idd}}^{2}$=0.0008~\AA$^{2}$ is not crucial --- nearly
identical results are obtained with a much larger value of
$\sigma_{\text{idd}}^{2}$=0.008~\AA$^{2}$. Despite the unavoidable
ambiguity, we nevertheless assume that the large difference between
$\sigma_{\text{idd}}^{2}$ and $\sigma_{\text{diff}}^{2}$ would be
maintained even if a more advanced modelling was employed.

Our interpretation that the polarization-dependence of the XAFS
damping in \crcu\ is a manifestation of the existence of very
different MSRD's depending on whether the pair-forming atoms belong to
the same basic molecular unit could be confirmed (or refuted) by
temperature-dependent measurements of the polarized spectra.  If XAFS
oscillations for \mm{\varepsilon \| z}\ are gradually restored when
the temperature decreases so that XAFS amplitudes become comparable
for \mm{\varepsilon \| z}\ and \mm{\varepsilon \perp z}, as suggested
by static-lattice calculations (see figures~\ref{fig-static}
and~\ref{fig-vibr}), it would be a strong argument in favour of our
interpretation.

Generally, our results suggest that whenever one observes different
damping rates of XAFS oscillations for different polarisations,
presence of atomic-pair-selective vibrations should be considered.
Special attention should be given in this respect to systems where the
interatomic distances can be split into different classes according to
their presumed stiffness.

The way we include the vibrations into XANES calculations (modifying
the free electron propagator) was used before but only to investigate
XANES in liquids close to critical conditions and relying on more
complicated equations that the simple substitution of
Eq.~(\ref{eqsubs}).\cite{HKF+99,HFN+03} In this work, we focus on a
crystal at room temperature and demonstrate that the simple formula
(\ref{eqsubs}) can be used to identify vibrational effects in x-ray
absorption spectra in the XANES and intermediate energy regions.  The
same approach could be applied to incorporate vibrations into other
spectroscopies that can be described within the real-space
multiple-scattering framework (such as, for example, of the x-ray
bremsstrahlung isochromat spectroscopy,\cite{SSV+89} x-ray
scattering\cite{FKF+04,KRS+11} or x-ray photoemission).\cite{Fuj09}


\section{Conclusions}

Atomic vibrations can be efficiently incorporated into x-ray
absorption spectra calculations based on the multiple-scattering
formalism by modifying the free-electron propagator via a damping
factor, $G^{pq}_{LL''} \rightarrow G^{pq}_{LL''} \exp
(-k^{2}\sigma_{pq}^{2})$.  If this procedure is used, the effects of
vibrations are treated within the same approximations which are
involved in the standard EXAFS formula while the multiple scattering
itself is treated exactly.

Creatinium tetrachlorocuprate \crcu\ crystal is an example of a system
where the effects of vibrations are significant in the XANES and
intermediate energy region (30--100~eV above the edge) even at room
temperature.  Fine structure oscillations in the experimental Cu $K$
edge XAS of \crCu\ are damped differently for the in-plane and
out-of-plane polarisations of the incoming x-rays.  The anisotropy of
the XAFS damping cannot be reproduced by the theory unless the
vibrations are taken into account in such a way that individual
molecular blocks within the \crCu\ crystal are treated as semi-rigid
entities while the mutual positions of these blocks are subject to
strong relative displacements.


\begin{acknowledgments}
This work was supported by the LD-COST~CZ program of the Ministry of
Education, Youth and Sport of the Czech Republic within the project
LD15097. Stimulating discussions with P.~Pattison are gratefully
acknowledged.
\end{acknowledgments}


\bibliography{liter_vibrations}


\end{document}